\documentclass{article}

\usepackage{PRIMEarxiv}

\usepackage[utf8]{inputenc} % allow utf-8 input
\usepackage[T1]{fontenc}    % use 8-bit T1 fonts
\usepackage{hyperref}       % hyperlinks
\usepackage{url}            % simple URL typesetting
\usepackage{booktabs}       % professional-quality tables
\usepackage{amsfonts}       % blackboard math symbols
\usepackage{nicefrac}       % compact symbols for 1/2, etc.
\usepackage{microtype}      % microtypography
\usepackage{lipsum}
\usepackage{fancyhdr}       % header
\usepackage{graphicx}       % graphics
\graphicspath{{media/}}     % organize your images and other figures under media/ folder
\usepackage{float}
\usepackage{amsmath}
\usepackage{subcaption}
\usepackage{multirow}  % para \multirow

\title{Pure and mixed Dicke state ansatz for equality and inequality constraints in variational quantum eigensolver}
\author{\href{https://orcid.org/0000-0001-9964-5973}{\includegraphics[scale=0.06]{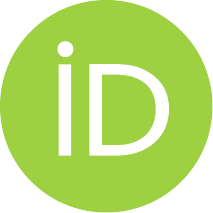}\hspace{1mm}J.V.S Scursulim}\\
  \texttt{josevictor.s.scursulim@gmail.com} \\
}

\begin{document}
\maketitle

\begin{abstract}
Combinatorial optimization can be addressed with quantum computing through variational quantum algorithms, but a central challenge in this approach is to design an ansatz expressive enough to explore the feasible subspace of the Hilbert space where the optimal solution lies. Another major challenge is tuning the Lagrange multipliers in penalty terms to enforce feasibility and guarantee solution quality. To address both challenges, we propose the first feasibility-preserving mixed Dicke state ansatz for Hamming weight constrained combinatorial optimization, extending the density matrix formalism to structurally encode equality and inequality constraints directly into the quantum circuit, thereby eliminating the need for penalty terms in the objective function. The proposed framework handles both constraint types, with the pure Dicke state ansatz recovered as a special case corresponding to equality constraints, and generalizes to multiple constraint groups via tensor products of individual pure or mixed Dicke states. We validate the proposed approach in the context of combinatorial portfolio optimization across three experimental scenarios with increasing constraint complexity, using the CMA-ES optimizer and comparing its performance against random search with replacement restricted to the feasible subspace. As the feasible search space grows, the proposed ansatz demonstrates a clear advantage over random search in terms of the number of objective function calls required to identify high-quality solutions. Hardware experiments on IBM NISQ processors confirm that noise mitigation and circuit transpilation optimizations remain open challenges for practical deployment. The framework is general and directly applicable to other combinatorial optimization problems with Hamming weight constraints.
\end{abstract}

% keywords can be removed

\section{Introduction}
\label{sec:intro}
Quantum computing is an emerging technology that offers potential computational advantages over classical computing for specific problem classes. This computing paradigm is based on the principles of quantum mechanics, where phenomena such as superposition, entanglement, relative phase, and interference are used as computational resources to achieve gains in algorithmic complexity \cite{deutsch1992rapid, bernstein1993quantum, shor1994algorithms, grover1996fast, harrow2009quantum}. In the near term, noisy quantum hardware devices, combined with error mitigation techniques, can yield computationally relevant results. In the long term, fault-tolerant quantum computers are expected to deliver significant advances in physics \cite{kim2023evidence, di2024quantum, farrell2024quantum, google2025observation, fischer2026dynamical}, chemistry \cite{kandala2017hardware, mcardle2020quantum, bauer2020quantum, von2021quantum, li2024hybrid, kaliakin2025implicit}, combinatorial optimization \cite{moll2018quantum, campbell2019applying, montanaro2020quantum, egger2021warm, abbas2024challenges, romero2025bias, kotil2025quantum, druagoi2026approximate}, machine learning \cite{biamonte2017quantum, liu2018quantum, cong2019quantum, havlivcek2019supervised, schuld2019quantum, huang2021power, cerezo2022challenges, glick2024covariant} and finance \cite{woerner2019quantum, egger2020quantum, stamatopoulos2020option, egger2020credit, scursulim2025multiclass}.

Combinatorial optimization arises in numerous problems of practical relevance across research, industry, and business, including the traveling salesman problem, portfolio optimization, tail assignment, facility location, Birkhoff decomposition, and others. These problems can be addressed with classical techniques such as mixed-integer programming and metaheuristics. However, as the number of variables and constraints scales, the computational cost of finding optimal solutions grows exponentially, often rendering exact methods intractable or forcing solvers to return suboptimal approximations within practical time limits.

Currently available quantum processors, while comprising hundreds to thousands of physical qubits, remain susceptible to noise, limiting their applicability to large-scale problems due to gate errors, decoherence, and limited qubit connectivity \cite{preskill2018quantum}. A promising near-term strategy combines error mitigation techniques \cite{kandala2018extending, giurgica2020digital, nation2021scalable, van2022model, filippov2023scalable, van2023probabilistic, kim2023scalable, gupta2024probabilistic, eddins2024lightcone, fuller2026improved}, optimization problem decomposition \cite{bass2021optimizing}, and variational quantum algorithms (VQAs) \cite{cerezo2021variational}. Two of the most studied VQAs are the Variational Quantum Eigensolver (VQE) \cite{peruzzo2014variational} and the Quantum Approximate Optimization Algorithm (QAOA) \cite{farhi2014quantum}, both of which alternate between parameterized quantum circuits and classical optimization routines. Even in the noise-free regime, VQAs face fundamental challenges that hinder the achievement of practical quantum advantage, including ansatz expressivity \cite{holmes2022connecting}, barren plateaus \cite{wang2021noise, larocca2025barren}, finite sampling noise \cite{scriva2024challenges}, and classical parameter optimization \cite{bittel2021training}.

In this work, we propose the first feasibility-preserving mixed Dicke state ansatz for Hamming weight constrained combinatorial optimization, extending the density matrix formalism and open quantum systems theory \cite{breuer2002theory, nielsen2010quantum} — previously explored in broader VQA contexts \cite{cerezo2022variational, zhao2023universal, shang2024hermitian, clemente2024mixed} — to the domain of structured combinatorial optimization with equality and inequality constraints. We provide a theoretical framework alongside experimental results that validate the proposed approach in the context of portfolio optimization. The framework is general and directly applicable to other combinatorial optimization problems with Hamming weight constraints, such as feature selection and ensemble selection \cite{mucke2023feature, leclerc2023financial}.

This manuscript is organized as follows. Section \ref{sec:comb_po} presents the combinatorial optimization problems addressed in this work. Section \ref{sec:dicke_vqe} discusses the theoretical aspects of the proposed ansatz. Section \ref{sec:methods} describes the experimental settings. Sections \ref{sec:results} and \ref{sec:discussion} present the experimental results and discuss the findings in light of the theoretical framework. Section \ref{sec:conclusion} provides the concluding remarks and directions for future work. Complementary discussions are presented in the Appendix.

\section{Combinatorial Portfolio Optimization}
\label{sec:comb_po}

Modern portfolio theory \cite{markowitz1952portfolio} provides a theoretical foundation for portfolio construction by balancing expected return and risk. In combinatorial portfolio optimization, the goal is to select a subset of assets that maximizes the expected return while minimizing risk. The mathematical formulation of this problem is given by the mean-variance model
\begin{eqnarray}
    \min_{x \ \in \ \{0,1\}^n}qx^T\Sigma x -(1-q)x^T\mu + r_f,
\end{eqnarray}
where $x$ is the binary decision variable vector, $\Sigma$ is the covariance matrix, $\mu$ is the vector of expected asset returns, $q \in [0,1]$ is the risk aversion factor that incorporates investor preferences, $r_f$ is the risk-free return, often approximated by the interest rate of the US Treasury. Since $r_f$ is a constant, it does not affect the optimal solution and is typically omitted in practice.

Beyond the objective function, constraints play a central role in portfolio optimization by encoding regulatory requirements, risk limits, and structural preferences. In combinatorial portfolio optimization, a natural constraint is to impose an upper bound, a lower bound, or an exact value on the Hamming weight of the decision variables, which in financial terms determines the number of assets selected to compose the portfolio. Such constraints are modeled mathematically as follows:
\begin{eqnarray}
    \sum_ix_i = b \label{eq_constr} \\
    \sum_ix_i \leq b \label{ineq_leq_constr} \\
    \sum_ix_i \geq b \label{ineq_geq_constr}.
\end{eqnarray}
Since $b \in \mathbb{N}$, strict inequality constraints can be rewritten as non-strict ones by replacing $b$ by $b+1$ and $b-1$ in equations (\ref{ineq_leq_constr}) and (\ref{ineq_geq_constr}) respectively. In practice, multiple constraints of the forms (\ref{eq_constr}), (\ref{ineq_leq_constr}), and (\ref{ineq_geq_constr}) can be combined to encode diversification requirements, generating more robust portfolios as highlighted in \cite{scursulim2025multiclass}.

\section{VQE with Pure and Mixed Dicke State Ansatz}
\label{sec:dicke_vqe}

Quantum computing addresses combinatorial optimization problems through the isomorphism between Quadratic Unconstrained Binary Optimization (QUBO) and Ising models, recasting the optimization problem as a ground state energy search. The optimization problem can be cast as a QUBO following the procedures described in \cite{glover2022quantum}, and through the change of variables $x_i=\frac{1-Z_i}{2}$ \cite{lucas2014ising}, the QUBO is converted into an Ising Hamiltonian that encodes the original problem. Quantum mechanics guarantees the existence of a ground state for physically valid Hamiltonians; however, finding this ground state exactly is generally intractable. Fortunately, the variational principle \cite{griffiths2018introduction} provides a mathematical framework to approximate the ground state energy: one proposes a parameterized trial wavefunction, commonly referred to as an ansatz, to evaluate the Hamiltonian expectation value. Inspired by the variational principle, the VQE algorithm combines a parameterized quantum circuit and a classical optimizer to minimize the Hamiltonian expectation value through the following objective function
\begin{eqnarray}
    \min_{\vec{\theta} \ \in \ \mathbb{R}^d} \langle\psi(\vec{\theta})\vert H \vert\psi(\vec{\theta})\rangle \label{vqe_pure_state}.
\end{eqnarray}
Parameterized quantum circuits are composed of rotation and entangling gates, but designing a circuit structure with sufficient expressivity to minimize (\ref{vqe_pure_state}) is generally a nontrivial task. However, structured optimization problems provide information that can be embedded directly into the ansatz: constraints of the forms (\ref{eq_constr}), (\ref{ineq_leq_constr}), and (\ref{ineq_geq_constr}) restrict the Hamming weight of the decision variables, suggesting that a quantum state designed to satisfy this property by construction is a natural ansatz choice — the Dicke state being the canonical example. The Dicke state originates from a physical model describing the interaction between light and matter \cite{garraway2011dicke}, and has since been studied in the context of quantum networks \cite{prevedel2009experimental}, quantum metrology \cite{toth2012multipartite}, quantum error correction \cite{ouyang2021permutation}, and quantum storage \cite{ouyang2021quantum}. Recent work has further proposed resource-efficient heralded schemes for their preparation in linear-optical platforms \cite{kang2026heralded}, broadening the prospects for practical implementation in quantum technologies.

In \cite{scursulim2025multiclass, wang2025variational}, a pure parameterized Dicke state is proposed as an ansatz for VQE to solve combinatorial portfolio optimization with the equality constraint (\ref{eq_constr}). This parameterized quantum state is given by
\begin{eqnarray}
    \vert D^n_k (\vec{\theta})\rangle = \sum_i\mathcal{P}_ic_i(\vec{\theta})\vert 0 \rangle^{\otimes(n-k)}\otimes\vert 1\rangle^k, \label{dicke_state}
\end{eqnarray}
where $c_i(\vec{\theta})$ is the probability amplitude and $\mathcal{P}_i$ denotes the $i$-th permutation operator acting on computational basis states with Hamming weight $k$. Substituting $\vert \psi(\vec{\theta})\rangle$ in (\ref{vqe_pure_state}) with (\ref{dicke_state}) using a Hamming weight $k$ that matches the equality constraint, the penalty term can be eliminated by setting its associated Lagrange multiplier to zero. Furthermore, (\ref{dicke_state}) defines a feasible search space whose size grows as $O(n^k)$. When $n \gg k$, this feasible subspace is exponentially smaller than the full search space of size $O(2^n)$, resulting in a significant reduction in the complexity of the optimization landscape.  

\begin{figure}[!htb]
  \centering
  \includegraphics[scale=0.4]{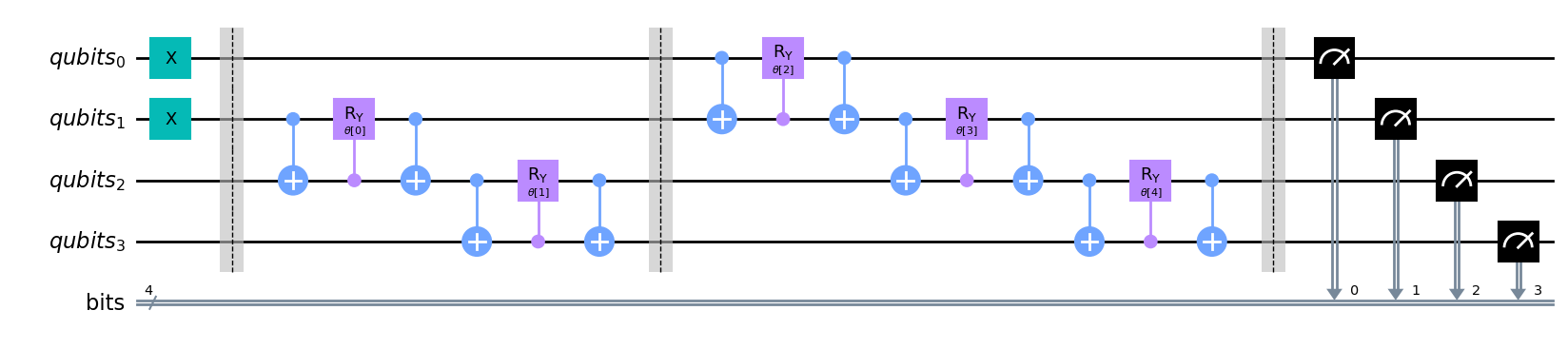}
  \caption{An example of a quantum circuit that generates a pure Dicke state ansatz for VQE. In this figure, the circuit shown prepares the parameterized quantum state $\vert D^4_2(\vec{\theta})\rangle$.}
  \label{fig:circuit_pure_dicke}
\end{figure}

Inequality constraints (\ref{ineq_leq_constr}) and (\ref{ineq_geq_constr}) are satisfied by any Dicke state $\vert D_k^n\rangle$ whose Hamming weight satisfies $k \leq b$ and $k \geq b$, respectively. A quantum state encoding a superposition over such Dicke states can be written as
\begin{eqnarray}
    \vert\phi(\vec{\alpha},\vec{\theta})\rangle = \sum_k\sqrt{p_k(\vec{\alpha})}\vert D^n_k (\vec{\theta})\rangle_A\otimes\vert k \rangle_B \label{multi_dicke},
\end{eqnarray}
where subsystem $B$, represented by $\vert k \rangle_B$ consists of ancilla qubits added to the quantum circuit to control the conditional preparation of each Dicke state.

\begin{figure}[!htb]
  \centering
  \includegraphics[scale=0.4]{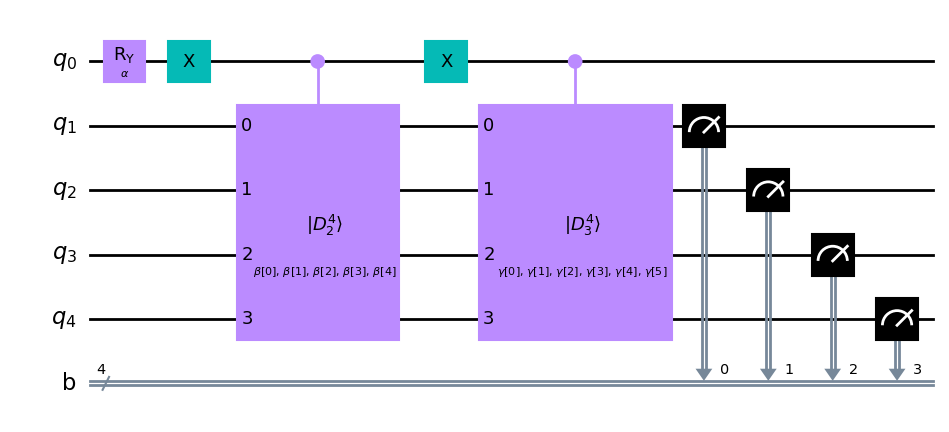}
  \caption{An example of a quantum circuit that creates a mixture of Dicke states as an ansatz for VQE. In this figure, the circuit shown creates a mixture of two quantum states, $\vert D^4_2(\vec{\beta})\rangle$ and $\vert D^4_3(\vec{\gamma})\rangle$, where the preparation probability of each state is regulated by the parameterized rotation gate $RY(\alpha)$. Note that the measurements only acts on the qubits where Dicke states are prepared, then they can be interpreted as a partial trace and providing us a mixture of quantum states.}
  \label{fig:circuit_mixture}
\end{figure}

Density matrices provide a general framework for describing both ensembles of quantum states and the dynamics of open quantum systems \cite{breuer2002theory, schlosshauer2007decoherence, nielsen2010quantum, weinberg2015lectures}. To construct an ansatz that generates an ensemble of Dicke states, we consider the density matrix obtained from the outer product of (\ref{multi_dicke}):
\begin{eqnarray}
    \rho_{AB}(\vec{\alpha},\vec{\theta}) &=& \vert\phi(\vec{\alpha},\vec{\theta})\rangle\langle\phi(\vec{\alpha},\vec{\theta})\vert \nonumber \\
    &=& \sum_{k,l}\sqrt{p_k(\vec{\alpha})}\sqrt{p_l(\vec{\alpha})}\vert D^n_k (\vec{\theta})\rangle\langle D^n_l (\vec{\theta})\vert\otimes\vert k \rangle\langle l \vert. \label{density_dicke}
\end{eqnarray}
Tracing the degrees of freedom of the subsystem $B$ from \ref{density_dicke}, we obtain the reduced density matrix
\begin{eqnarray}
    \sigma(\vec{\alpha},\vec{\theta}) &=& \textrm{tr}_B(\rho_{AB}) \nonumber \\
    &=& \sum_k p_k(\vec{\alpha})\vert D^n_k (\vec{\theta})\rangle\langle D^n_k (\vec{\theta})\vert, \label{mixture_dicke}
\end{eqnarray}
which describes a mixture of Dicke states with different Hamming weight values, where $p_k(\vec{\alpha})$ dictates the probability of preparing $\vert D^n_k (\vec{\theta})\rangle$. Figure \ref{fig:circuit_mixture} shows an example of a quantum circuit that produces the mixture described by (\ref{mixture_dicke}). Note that the ancilla qubit represents subsystem $B$ in (\ref{multi_dicke}), and that the measurement gates implement the partial trace operation.

In the density matrix formalism, the expectation value of an operator $O$ is given by $\langle O \rangle = \textrm{tr}(\rho O)$ \cite{breuer2002theory, schlosshauer2007decoherence, nielsen2010quantum, weinberg2015lectures}, where $\rho$ is the density matrix of the system. Accordingly, the VQE objective function generalizes to
\begin{eqnarray}
    \min_{\vec{\alpha},\vec{\theta} \ \in \ \mathbb{R}^d}\textrm{tr}(\sigma(\vec{\alpha},\vec{\theta})H). \label{vqe_mixed_state}
\end{eqnarray}
Note that when $\sigma = \vert \psi(\vec{\theta})\rangle\langle\psi(\vec{\theta})\vert$ is a pure state, (\ref{vqe_mixed_state}) reduces to (\ref{vqe_pure_state}). For combinatorial optimization problems mapped into an Ising model, the Hamiltonian is diagonal, $H = \sum_{x \in \{0,1\}^n} E_x |x\rangle\langle x|$, the equation (\ref{vqe_mixed_state}) reduces to 
\begin{eqnarray}
    \min_{\vec{\alpha},\vec{\theta} \ \in \ \mathbb{R}^d}\sum_m\sigma_{mm}(\vec{\alpha},\vec{\theta})E_m,
\end{eqnarray}
where $\sigma_{mm}(\vec{\alpha},\vec{\theta})$ are the diagonal terms of the density matrix $\sigma(\vec{\alpha},\vec{\theta})$, corresponding to the probability distribution over the computational basis states obtained from measurements on the quantum circuit. For a detailed derivation of (\ref{vqe_mixed_state}), the reader is referred to Appendix \ref{appendix:a}.

The framework introduced above naturally extends to optimization problems with multiple Hamming weight constraints by exploiting the tensor product structure of composite quantum systems. Consider $C$
groups of $n_c$ decision variables, where group $c$ is subject to a constraint of the form (\ref{eq_constr}), (\ref{ineq_leq_constr}), or (\ref{ineq_geq_constr}) with bound $b_c$. The global ansatz is then given by the tensor product of the individual mixtures
\begin{eqnarray}
    \sigma(\vec{\alpha},\vec{\theta}) = \bigotimes_{c=1}^{C} \sigma_c(\vec{\alpha}_c, \vec{\theta}_c) = \bigotimes_{c=1}^{C} \sum_{k_c} p_{k_c}(\vec{\alpha}_c) \vert D^{n_c}_{k_c}(\vec{\theta}_c)\rangle\langle D^{n_c}_{k_c}(\vec{\theta}_c)\vert,
\end{eqnarray}
where each subsystem has its own parameter vectors $\vec{\alpha}_c$ and $\vec{\theta}_c$. This construction satisfies all constraints simultaneously by design, without requiring penalty terms. Furthermore, it subsumes the pure state ansatz of \cite{scursulim2025multiclass} as a special case, recovered when each mixture $\sigma_c$ collapses to a single Dicke state with fixed Hamming weight $k_c = b_c$. This formulation also accommodates mixed constraint types within the same problem: subsystems subject to equality constraints (\ref{eq_constr}) are represented by pure Dicke states with fixed Hamming weight $k_c = b_c$, while subsystems subject to inequality constraints (\ref{ineq_leq_constr}) or (\ref{ineq_geq_constr}) are represented by genuine mixtures over the feasible range of Hamming weights.

The number of variational parameters in the ansatz is directly related to the number of constraints, the number of qubits, and the Hamming weight bounds. The total parameter count of the quantum circuit is given by
\begin{eqnarray}
    n_p = \sum_{i=1}^{s}\sum_{k=k_{min,i}}^{k_{max,i}}kn_{i}-\frac{k(k+1)}{2} + \sum_{i=1}^s(k_{max,i}-k_{min,i}),
\end{eqnarray}
where $s$ is the number of constraint groups, and $k_{min,i}$ and $k_{max,i}$ are the Hamming weight lower and upper bounds for the $i$-th group. Note that for equality constraints, where $k_{min,i} = k_{max,i}$, the ancilla term vanishes and the expression reduces to the parameter count of a single pure Dicke state ansatz for each constraint group, consistent with \cite{scursulim2025multiclass}.

Although the experimental validation in this work focuses on portfolio optimization, the proposed ansatz is applicable to any combinatorial optimization problem whose feasibility conditions can be expressed as equality or inequality constraints on the Hamming weight of the decision variables, such as feature selection \cite{mucke2023feature} and ensemble selection \cite{leclerc2023financial}

\section{Methods}
\label{sec:methods}

To evaluate the effectiveness of the proposed ansatz, we design three experimental scenarios. In all scenarios, the case $k = 0$, corresponding to an empty portfolio with no assets selected, is excluded from the feasible set by setting the lower bound of each mixture to $k_{min} \geq 1$, as it has no financial relevance. The size of the feasible search space is computed by $\sum_k C_{n,k}$ single-group scenarios, where $C_{n,k} = n!/k!(n-k)!$ is the binomial coefficient, and as $\prod_i\sum_{k_i} C_{n_i,k_i}$ for multi-group scenarios involving tensor product ansatze.

Scenario I consists of a portfolio optimization with 11 assets subject to the inequality constraint $k \leq 4$, requiring two ancilla qubits to prepare the mixture
\begin{eqnarray}
    \sigma_I(\vec{\alpha},\vec{\theta}) = \sum_{k=1}^4 p_k(\vec{\alpha})\vert D^{11}_k (\vec{\theta})\rangle\langle D_k^{11}(\vec{\theta})\vert. \label{ansatzI}
\end{eqnarray}
The feasible search space for this scenario has size $\vert \mathcal{F}_I\vert = \sum_{k=1}^4C_{11,k}=561$.
Scenario II extends the previous one by introducing two simultaneous inequality constraints, $k \geq 3$ and $k \leq 6$, so that the ansatz becomes
\begin{eqnarray}
    \sigma_{II}(\vec{\alpha},\vec{\theta}) = \sum_{k=3}^6 p_k(\vec{\alpha})\vert D^{11}_k (\vec{\theta})\rangle\langle D_k^{11}(\vec{\theta})\vert. \label{ansatzII}
\end{eqnarray}
The feasible search space for this scenario has size $\vert \mathcal{F}_{II}\vert = \sum_{k=3}^6C_{11,k}=1419$.
Scenario III was designed to validate the tensor product construction and to test a setting combining equality and inequality constraints. We construct a portfolio optimization instance by randomly selecting assets from 4 sectors of the S\&P 500 index — Energy, Financial Services, Real Estate, and Basic Materials — with 5 assets per sector. The constraints were defined as follows: an equality constraint $k_E = 3$ for Energy, two inequality constraints $1 \leq k_{FS} \leq 2$ for Financial Services, an equality constraint $k_{RE} = 2$ for Real Estate, and two inequality constraints $1 \leq k_{BM} \leq 4$ for Basic Materials. To prepare the ansatz, 3 ancilla qubits were added, yielding the following tensor product mixture
\begin{eqnarray}
    \sigma_{III}(\vec{\alpha},\vec{\theta}) = \sigma_E \otimes \sigma_{RE} \otimes \sigma_{FS} \otimes \sigma_{BM}. \label{ansatzIII}
\end{eqnarray}
where each subsystem density matrix is defined as
\begin{eqnarray}
    \sigma_E = \vert D_3^5(\vec{\theta})\rangle\langle D_3^5(\vec{\theta})\vert, \quad 
    \sigma_{RE} = \vert D_2^5(\vec{\theta})\rangle\langle D_2^5(\vec{\theta})\vert,
\end{eqnarray}
\begin{eqnarray}
    \sigma_{FS} = \sum_{k_{FS}=1}^{2} p_{k_{FS}}(\vec{\alpha})\vert D_{k_{FS}}^5(\vec{\theta})\rangle\langle D_{k_{FS}}^5(\vec{\theta})\vert, \quad
    \sigma_{BM} = \sum_{k_{BM}=1}^{4} p_{k_{BM}}(\vec{\alpha})\vert D_{k_{BM}}^5(\vec{\theta})\rangle\langle D_{k_{BM}}^5(\vec{\theta})\vert.
\end{eqnarray}
The feasible search space for this scenario has size $\vert \mathcal{F}_{III}\vert = \prod_i\sum_{k_i}C_{n_i,k_i}=45000$.

The CMA-ES optimizer \cite{grayver2016exploring} was selected as the primary optimizer based on its performance in \cite{scursulim2025multiclass} and because it is gradient-free, as further discussed in Appendix \ref{appendix:b}. Experiments were also conducted with COBYLA \cite{powell1994direct} for comparison purposes; however, given its substantially lower success rate in preliminary analysis, the main results focus on CMA-ES. In all experiments, we evaluated $2^i$ shots for $i \in \{0, 1, \ldots, 12\}$ and a maximum number of iterations of $10^j$ for $j \in \{0, 1, 2, 3\}$. For each scenario, 100 independent experiments were conducted for each combination of shots and iterations, each initialized from a distinct randomly generated initial point, with the random seed fixed to 42 for reproducibility.

All simulations were performed on a personal workstation running Ubuntu 24.04.4 LTS, equipped with an AMD Ryzen 9 9950X3D processor (16 cores, 32 threads) and 64 GB of DDR5 RAM clocked at 6400 MHz. The first two experimental scenarios were simulated using the CPU, while the third scenario, due to its larger scale, required GPU acceleration and was executed on an NVIDIA GeForce RTX 4080 Super with 16 GB of VRAM. All code was developed in Python 3.11, making use of the following packages: NumPy, Matplotlib, Seaborn, Pandas, SciPy, pylatexenc, Qiskit \cite{aleksandrowicz2019qiskit}, Qiskit IBM Runtime, Qiskit Aer GPU (CUDA 11), ipykernel, Plotly, Qiskit Addon Optimization Mapper, CPLEX, Gurobi, CMA-ES \cite{grayver2016exploring}, Optuna, yfinance, tqdm, nbformat, and Qiskit Experiments. The complete computational environment, including package versions and Python interpreter, is specified in a Pipfile available in the project repository (see Code Availability statement).

\section{Results}
\label{sec:results}

This section presents the simulation results. As mentioned in Section \ref{sec:methods}, experiments were performed with both CMA-ES and COBYLA. However, due to COBYLA's constraints on setting a maximum number of iterations and its substantially lower success rate in preliminary analysis, we focus on CMA-ES results. Our goal is not to determine which optimizer performs best, but to validate the proposed framework. Interestingly, in the few cases where COBYLA successfully identified the optimal solution, the resulting distribution was nearly concentrated on the target bitstring, suggesting that its primary limitation in this setting is robustness to finite sampling noise rather than convergence quality.

\begin{figure}[!htbp]
    \centering
    \begin{subfigure}[!hb]{0.45\textwidth}
        \centering
        \includegraphics[width=\textwidth]{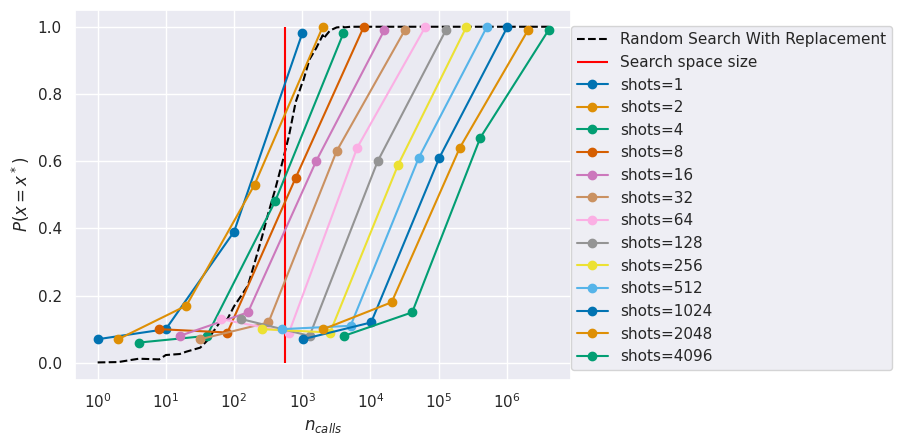}
        \caption{Scenario I}
    \end{subfigure}
    \hfill
    \begin{subfigure}[!hb]{0.45\textwidth}
        \centering
        \includegraphics[width=\textwidth]{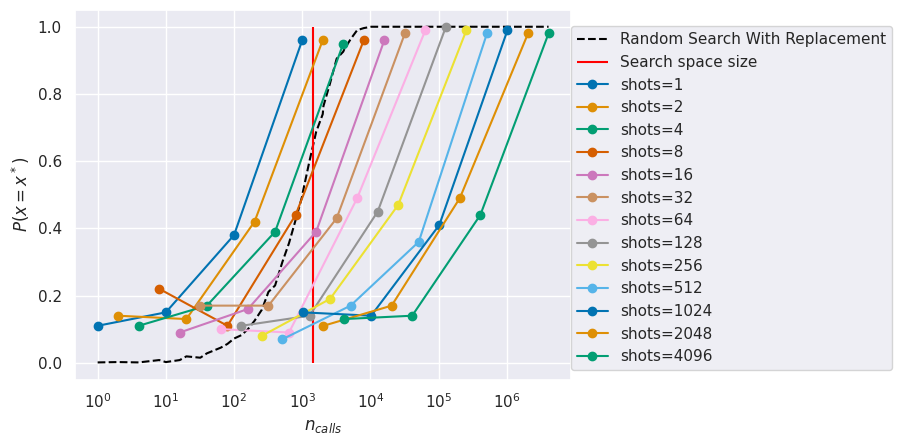}
        \caption{Scenario II}
    \end{subfigure}
    \hfill
    \begin{subfigure}[!hb]{0.45\textwidth}
        \centering
        \includegraphics[width=\textwidth]{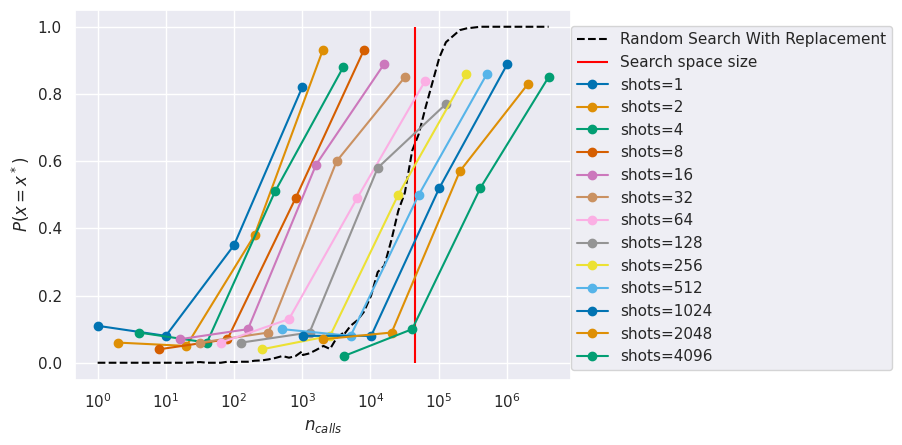}
        \caption{Scenario III}
    \end{subfigure}
    \hfill
    \caption{Probability of finding the optimal solution $P(x = x^*)$ as a function of the number of objective function calls $n_{calls}$ for Scenarios I (left, $|\mathcal{F}_I| = 561$), II (center, $|\mathcal{F}_{II}| = 1419$), and III (right, $|\mathcal{F}_{III}| = 45000$), using the CMA-ES optimizer. Each curve corresponds to a different number of measurement shots, and each point represents the empirical success probability estimated over 100 independent runs, where success is defined as the optimal bitstring $x^*$ having non-zero sampling probability after training with the corresponding number of objective function calls. The dashed black curve shows the expected performance of random search with replacement restricted to the feasible search space, and the vertical red line indicates the feasible search space size $|\mathcal{F}|$. Curves crossing the red line to the left indicate configurations where the proposed ansatz outperforms random search in terms of the number of objective function calls. The advantage over random search becomes more pronounced as the feasible search space size increases from Scenario I to Scenario III.}
    \label{fig:random_search}
\end{figure}

\begin{figure}[!htbp]
    \centering
    \begin{subfigure}[!hb]{0.3\textwidth}
        \centering
        \includegraphics[width=\textwidth]{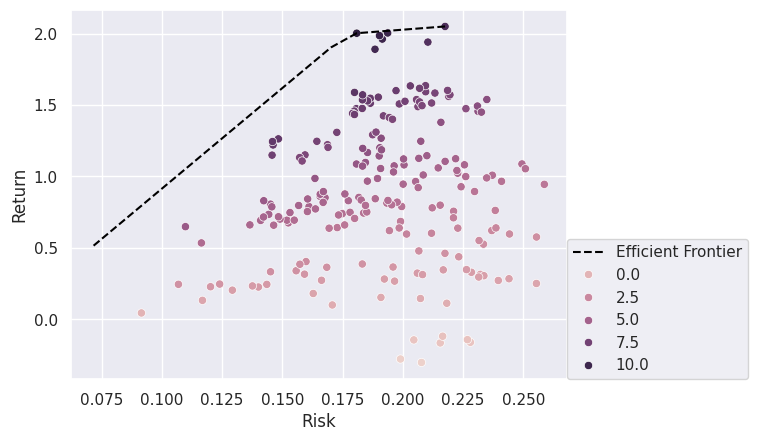}
        \caption{Scenario I}
    \end{subfigure}
    \hfill
    \begin{subfigure}[!hb]{0.3\textwidth}
        \centering
        \includegraphics[width=\textwidth]{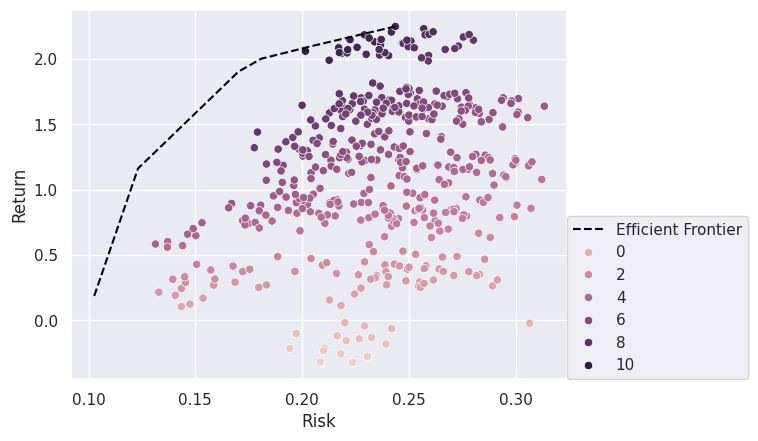}
        \caption{Scenario II}
    \end{subfigure}
    \hfill
    \begin{subfigure}[!hb]{0.3\textwidth}
        \centering
        \includegraphics[width=\textwidth]{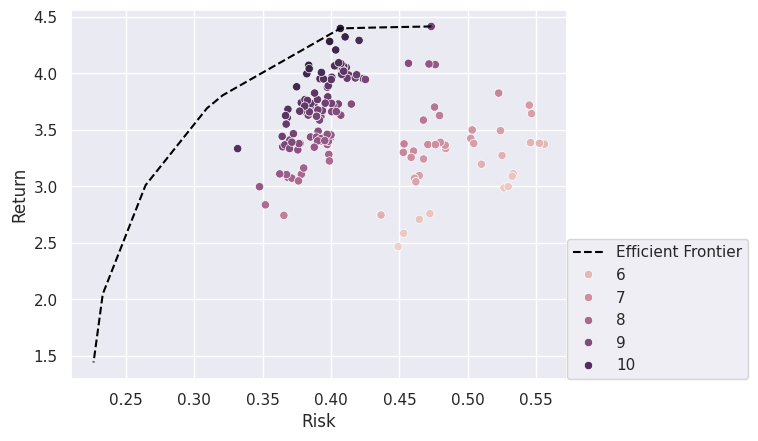}
        \caption{Scenario III}
    \end{subfigure}
    \hfill
    \caption{Discrete efficient frontier for Scenarios I (left), II (center), and III (right), showing portfolios sampled from the experiment in which the target bitstring $x^*$ was observed with the highest sampling probability after training, selected among 100 independent runs per scenario. Each point represents a feasible portfolio, colored by Sharpe ratio. The dashed line indicates the discrete efficient frontier computed via classical mean-variance optimization. In all three scenarios, the proposed ansatz successfully identifies portfolios on or near the efficient frontier.}
    \label{fig:efficient_frontier}
\end{figure}

Figures \ref{fig:random_search} and \ref{fig:efficient_frontier} present the main simulation results. Figure \ref{fig:random_search} shows the probability of finding the optimal solution as a function of the number of objective function calls for all three scenarios, compared against random search with replacement restricted to the feasible subspace. Figure \ref{fig:efficient_frontier} shows the portfolios sampled in the best performing experiment for each scenario, positioned relative to the classical efficient frontier.

Figure \ref{fig:random_search} shows the probability of finding the optimal solution $P(x = x^*)$ as a function of the number of objective function calls $n_{calls}$, comparing the VQE-Dicke framework against random search with replacement restricted to the feasible subspace. As the feasible search space size increases from Scenario I to Scenario III, the advantage of VQE-Dicke over random search becomes more pronounced, as the optimization process progressively concentrates probability mass toward the optimal solution. This behavior is consistent with the theoretical argument that the proposed ansatz is most effective in the regime where the feasible search space is large enough to make exhaustive random sampling impractical. All CMA-ES and COBYLA results for each scenario and experimental setting described in Section \ref{sec:methods} are available in the project repository (see Data Availability statement).

Figure \ref{fig:efficient_frontier} illustrates how the bitstrings sampled from the optimized distribution can be used to analyze the quality of feasible solutions in the context of portfolio optimization. These results highlight that even when the optimizer fails to drive the reduced density matrix to fully concentrate on the optimal bitstring $x^*$, the sampled bitstrings still provide a set of high-quality feasible solutions that can be further analyzed or used as warm starts for classical post-processing.

\section{Discussion}
\label{sec:discussion}

The results presented in Section \ref{sec:results} demonstrate that the VQE-Dicke framework successfully identifies optimal or near-optimal portfolio solutions across all three experimental scenarios, with a clear advantage over random search with replacement as the feasible search space grows. In this section, we discuss the theoretical and practical implications of these findings. We first analyze the computational complexity of the proposed framework and the reduction of the feasible search space afforded by the ansatz construction. We then discuss the ideal convergence behavior of the optimization process and the practical strategy for extracting high-quality solutions from the sampled distribution. Finally, we address the impact of hardware noise on the proposed ansatz, with particular attention to bit-flip and readout errors, and discuss the robustness properties of the mixed Dicke state ansatz in the presence of such noise.

The computational complexity of VQE scales as $\mathcal{O}(N_{iter}\times N_{shots} \times N_{gcp})$, where $N_{gcp}$ is the number of groups of commuting Pauli operators required to estimate the Hamiltonian expectation value. For combinatorial optimization problems mapped into an Ising model, the Hamiltonian is diagonal and all its terms commute, yielding $N_{gcp}=1$ and thereby reducing the complexity to $\mathcal{O}(N_{iter}\times N_{shots})$. Furthermore, the ansatze defined by (\ref{dicke_state}) and (\ref{mixture_dicke}) satisfy constraints (\ref{eq_constr}), (\ref{ineq_leq_constr}), and (\ref{ineq_geq_constr}) by construction, reducing the search space from $\mathcal{O}(2^n)$, which includes infeasible solutions, to the feasible subspace of size $\mathcal{O}(n^k)$, which is exponentially smaller than $\mathcal{O}(2^n)$ in the regime $n \gg k$. For problems with multiple constraints, the feasible search space is dominated by $\mathcal{O}(n^{k_{max}})$, which remains exponentially smaller than $\mathcal{O}(2^n)$, under the assumption $n \gg k_{max}$.

Considering that $\vec{\alpha}^*$ and $\vec{\theta}^*$ are the optimal parameter vectors and $x^*$ is the optimal solution, the ideal behavior of the ansatz after optimization is for the reduced density matrix to converge to
\begin{eqnarray}
    \sigma(\vec{\alpha}^*,\vec{\theta}^*) = |x^*\rangle\langle x^*|,
\end{eqnarray}
corresponding to a distribution concentrated entirely on the optimal bitstring $x^*$. In the continuous analogy, this is equivalent to a Dirac delta distribution $\delta(x-x^*)$ centered at $x^*$, which in the discrete setting becomes a Kronecker delta $\delta_{x,x^*}$ over the feasible search space. Achieving this ideal behavior, however, may require a large number of optimizer iterations and circuit evaluations, potentially exceeding the feasible search space size $\vert \mathcal{F}\vert$ and performing worse than random search with replacement in terms of the number of objective function calls, as discussed in \cite{scriva2024challenges}. In practice, a more pragmatic strategy is to sample the optimized circuit and analyze the most frequently observed bitstrings, selecting the one that yields the lowest objective function value. This approach leverages the fact that, even when $\sigma(\vec{\alpha}^*,\vec{\theta}^*)$ does not fully collapse to $|x^*\rangle\langle x^*|$, the optimization process tends to concentrate probability mass around high-quality solutions, making the most frequent bitstrings strong candidates for the optimal or near-optimal solution. As illustrated in Figure \ref{fig:efficient_frontier}, the bitstrings sampled from the optimized distribution can be mapped onto the risk-return space and compared against the classical efficient frontier, providing a practical tool for portfolio analysis even in cases where perfect convergence is not achieved.

Finally, it is important to discuss the impact of hardware noise on the proposed ansatz when considering its execution on real quantum devices. The results obtained through real quantum devices will be discussed in more detail in the Appendix \ref{appendix:c}. Quantum states defined by (\ref{dicke_state}) and (\ref{mixture_dicke}), due to their characteristic Hamming weight structure, are particularly sensitive to bit-flip and readout errors, since these errors can alter the Hamming weight of each state in the superposition. The bit-flip channel acting on a single-qubit density matrix $\rho$ is modeled by \cite{breuer2002theory, schlosshauer2007decoherence, nielsen2010quantum}
\begin{eqnarray}
    \mathcal{E}(\rho)=(1-p)\rho + pX\rho X.
\end{eqnarray}
It is worth noting that the proposed ansatz circuit relies heavily on two-qubit gates for the conditional preparation of each Dicke state block, and two-qubit gate errors are typically the dominant source of noise in superconducting quantum hardware \cite{kim2023evidence, krantz2019quantum}. A more realistic error model would therefore consider a two-qubit bit-flip channel acting on each pair of interacting qubits \cite{breuer2002theory, schlosshauer2007decoherence, nielsen2010quantum}
\begin{eqnarray}
    \mathcal{E}_{ij}(\rho) = (1-p)^2\rho + p(1-p)(X_i\rho X_i + X_j\rho X_j) + p^2 X_iX_j\rho X_iX_j,
\end{eqnarray}
where $p$ is the error probability per two-qubit gate, and $i$, $j$ denote the qubit indices involved in each gate. The evolution of the parameterized mixed state (\ref{mixture_dicke}) under two-qubit bit-flip channels acting on all interacting qubit pairs in the circuit is given by
\begin{eqnarray}
    \sigma'(\vec{\alpha},\vec{\theta}) = \bigotimes_{(i,j) \in \mathcal{G}} \mathcal{E}_{ij}(\sigma(\vec{\alpha},\vec{\theta})),
\end{eqnarray}
where $\mathcal{G}$ denotes the set of interacting qubit pairs defined by the circuit connectivity, and $\mathcal{E}_{ij}$ denotes the two-qubit bit-flip channel acting on qubits $i$ and $j$. In the case of (\ref{dicke_state}), a single bit-flip is sufficient to corrupt the Hamming weight and displace the state outside the feasible set. For (\ref{mixture_dicke}), however, only states at the boundary of the feasible range, i.e. those with $k = k_{min}$ or $k=k_{max}$, can be displaced outside the feasible set by a single bit-flip, whereas multiple simultaneous bit-flips can compromise interior states as well. This implies that (\ref{mixture_dicke}) exhibits greater robustness to single bit-flip errors compared to (\ref{dicke_state}), as the mixture over multiple Hamming weight values preserves a larger portion of the feasible search space. Nevertheless, both error types impact the objective functions (\ref{vqe_pure_state}) and (\ref{vqe_mixed_state}), potentially degrading the optimization process. It is therefore important to employ error mitigation techniques targeting readout and bit-flip errors when executing this ansatz on quantum hardware, so as to preserve the Hamming weight structure of the prepared states.

\section{Conclusion}
\label{sec:conclusion}

In this work, we introduced a variational quantum optimization framework that leverages pure and mixed Dicke states to structurally encode both equality and inequality constraints directly into the quantum circuit ansatz. By operating within the density matrix formalism and applying partial traces via ancilla measurements, we eliminated the need for penalty terms and Lagrange multiplier tuning in the objective function.

Our experimental validation on combinatorial portfolio optimization across three distinct constraint settings — including a complex multi-sector setup utilizing tensor product mixtures — demonstrated that the proposed ansatz can significantly outperform random sampling restricted to the feasible search space in terms of objective function calls. Furthermore, optimization runs successfully mapped assets onto or near the classical efficient frontier. Hardware implementations on IBM NISQ processors highlighted the inherent trade-offs: while the ansatz exhibits greater theoretical robustness to single bit-flip errors affecting boundary states, it remains sensitive to cumulative two-qubit gate noise, yielding a relative error of approximately $50\%$ across QPUs.

Future work will investigate the integration of advanced error mitigation techniques, such as zero-noise extrapolation or probabilistic error cancellation, combined with circuit transpilation optimizations, to reduce the performance gap between simulation and hardware execution. The proposed framework is general and directly applicable to other combinatorial optimization problems with Hamming weight constraints, such as feature selection and ensemble selection, opening promising directions for further experimental validation.

% \section*{Acknowledgments}
% This was supported in part by......

\section*{Disclaimer}
This work is purely academic. The portfolio optimization problem is used solely as a benchmark for combinatorial optimization methods and does not constitute financial advice or investment recommendations.

\section*{Data availability}

The data used in this study are publicly available through Yahoo Finance API and can also be obtained using scripts available in the following GitHub repository: \url{https://github.com/jvscursulim/quantum-optimization-dicke-states}. The dataset covers the period 05/05/2025 to 05/04/2025 and asset identifiers have been anonymized.

\section*{Code availability}

All code developed during this research is publicly available in the following GitHub repository: \url{https://github.com/jvscursulim/quantum-optimization-dicke-states}.

%Bibliography
\bibliographystyle{unsrt}  
\bibliography{references}

\appendix

\section{Derivation of the Mixed-State VQE Objective Function}
\label{appendix:a}

% \begin{eqnarray}
%     \langle H \rangle &=& \textrm{tr}(\sigma(\vec{\alpha},\vec{\theta}) H); \  H = \sum_{x \ \in \ \{0,1\}^n} E_x\vert x \rangle\langle x \vert\nonumber\\
%     &=& \textrm{tr}\left(\sum_{x \ \in \ \{0,1\}^n}\sum_k p_k(\vec{\alpha})E_x\vert D^n_k (\vec{\theta})\rangle\langle D^n_k (\vec{\theta})\vert x \rangle\langle x \vert\right); \vert D^n_k (\vec{\theta})\rangle = \sum_{{y \ \in \ \{0,1\}^n}, \ {\textrm{hw}(y) = k}}c_y(\vec{\theta})\vert y \rangle  \nonumber\\
%     &=& \textrm{tr}\left(\sum_k\sum_{x,  \ \in \ \{0,1\}^n}\sum_{{y \ \in \ \{0,1\}^n}, \ {\textrm{hw}(y) = k}}p_k(\vec{\alpha})\vert c_y(\vec{\theta})\vert^2\vert y \rangle\langle y \vert x \rangle\langle x \vert\right); \ \delta_{xy} = \langle y \vert x \rangle \nonumber \\
%     &=& \textrm{tr}\left(\sum_k p_k(\vec{\alpha})E_k\vert D^n_k (\vec{\theta})\rangle\langle D^n_k (\vec{\theta})\vert\right) \nonumber \\
%     &=& \sum_k p_k(\vec{\alpha})E_k\textrm{tr}\left(\vert D^n_k (\vec{\theta})\rangle\langle D^n_k (\vec{\theta})\vert\right) \nonumber\\
%     &=& \sum_k \sum_i \vert c_{i,k}(\vec{\theta}) \vert^2 p_k(\vec{\alpha})E_k \nonumber \\
%     &=& \sum_k \sigma_{kk}(\vec{\alpha}, \vec{\theta})E_k
% \end{eqnarray}

We derive here the simplified form of the VQE objective function \ref{vqe_mixed_state} for combinatorial optimization problems whose Hamiltonian is diagonal in the computational basis, i.e. $H = \sum_{x \in \{0,1\}^n} E_x |x\rangle\langle x|$. The derivation exploits the linearity of the trace operation and the structure of the mixed Dicke state ansatz \ref{mixture_dicke}.
\begin{eqnarray}
\langle H \rangle &=& \textrm{tr}(\sigma(\vec{\alpha},\vec{\theta}) H) \nonumber\\
&=& \textrm{tr}\left(\sum_k p_k(\vec{\alpha}) 
    |D^n_k(\vec{\theta})\rangle\langle D^n_k(\vec{\theta})| 
    \sum_{x \in \{0,1\}^n} E_x |x\rangle\langle x|\right) \nonumber\\
&=& \sum_k p_k(\vec{\alpha}) \sum_{x \in \{0,1\}^n} E_x\, 
    \textrm{tr}\left(|D^n_k(\vec{\theta})\rangle
    \langle D^n_k(\vec{\theta})|x\rangle\langle x|\right) \nonumber\\
&=& \sum_k p_k(\vec{\alpha}) \sum_{x \in \{0,1\}^n} E_x\, 
    |\langle x | D^n_k(\vec{\theta})\rangle|^2 \nonumber\\
&=& \sum_k \sum_{\substack{x \in \{0,1\}^n \\ 
    \textrm{hw}(x)=k}} p_k(\vec{\alpha})|c_x(\vec{\theta})|^2\,E_x \nonumber\\
&=& \sum_{m} \sigma_{mm}(\vec{\alpha},\vec{\theta})\, E_m
\end{eqnarray}
The final expression shows that the objective function reduces to a weighted sum of energies $E_x$ with weights given by the diagonal elements $\sigma_{mm}(\vec{\alpha}, \vec{\theta}) = p_{\textrm{hw}(m)}(\vec{\alpha})|c_m(\vec{\theta})|^2$, which correspond to the measurement outcome probabilities extracted directly from the quantum circuit.

\section{Effects of Finite Sampling on the Variational Optimization Landscape}
\label{appendix:b}

\begin{figure}[!htb]
  \centering
  \includegraphics[scale=0.35]{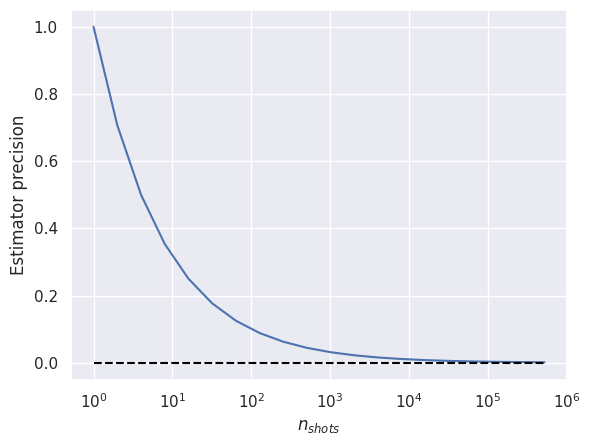}
  \caption{Estimator standard deviation $\sigma_{\langle O\rangle}$ as a function of $n_{shots}$ illustrating the $1/\sqrt{n_{shots}}$ decay. The dashed line indicates the noiseless statevector limit, recovered as $n_{shots} \rightarrow \infty$.}
  \label{fig:estimator_precision}
\end{figure}

In this appendix, we examine the effects of finite sampling on the variational optimization landscape, since in practice we have access only to samples from the probability distribution associated with the quantum state prepared by the circuit. The standard deviation of the expectation value of an operator $O$ follows 
\begin{eqnarray}
    \sigma_{\langle O\rangle} \propto \frac{1}{\sqrt{n_{shots}}}. \label{estimator_precision}
\end{eqnarray}
In the noiseless limit, this quantity satisfies
\begin{eqnarray}
    \lim_{n_{shots} \rightarrow \infty}\sigma_{\langle O\rangle} = 0,
\end{eqnarray}
corresponding to the regime where the complete probability distribution is recovered. Figure \ref{fig:estimator_precision} illustrates the behavior of \ref{estimator_precision} as the number of shots increases, where the dashed line indicates the noiseless limit $\sigma_{\langle O\rangle} = 0$.

\begin{figure}[!htbp]
    \centering
    \begin{subfigure}[!hb]{0.19\textwidth}
        \centering
        \includegraphics[width=\textwidth]{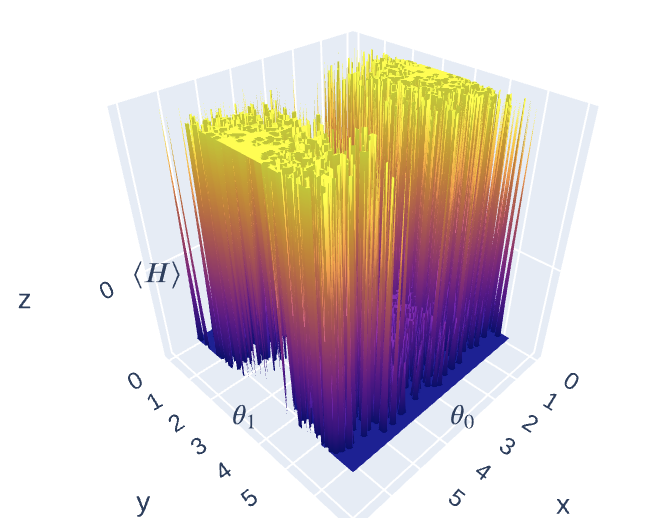}
        \caption{$n_{shots}=1$}
    \end{subfigure}
    \hfill
    \begin{subfigure}[!hb]{0.19\textwidth}
        \centering
        \includegraphics[width=\textwidth]{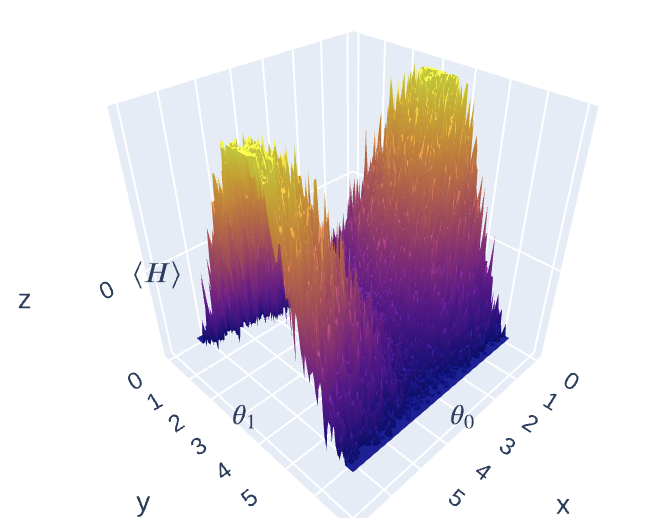}
        \caption{$n_{shots}=16$}
    \end{subfigure}
    \hfill
    \begin{subfigure}[!hb]{0.19\textwidth}
        \centering
        \includegraphics[width=\textwidth]{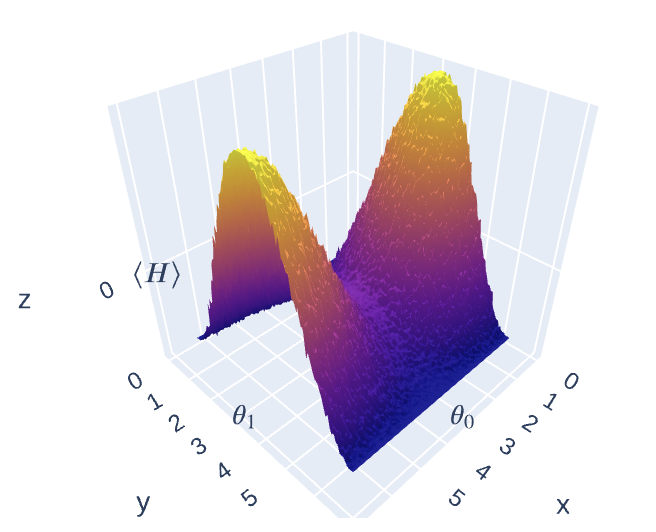}
        \caption{$n_{shots}=256$}
    \end{subfigure}
    \hfill
    \begin{subfigure}[!hb]{0.19\textwidth}
        \centering
        \includegraphics[width=\textwidth]{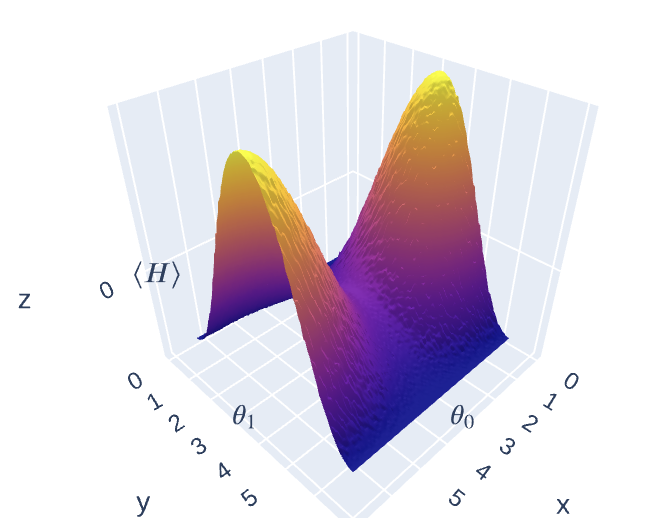}
        \caption{$n_{shots}=1024$}
    \end{subfigure}
    \hfill
    \begin{subfigure}[!hb]{0.19\textwidth}
        \centering
        \includegraphics[width=\textwidth]{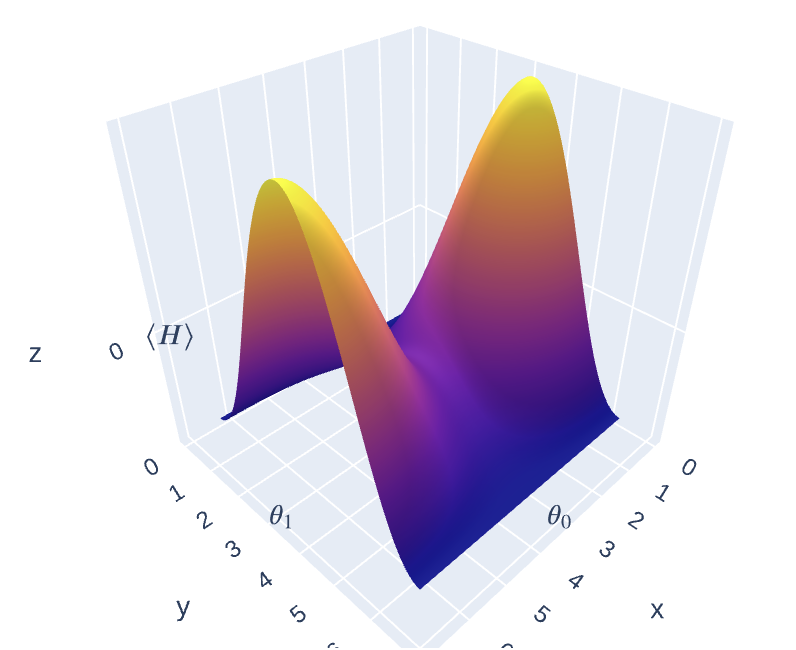}
        \caption{$n_{shots} \rightarrow \infty$}
    \end{subfigure}
    \caption{Energy landscape $\langle H \rangle$ as a function of the variational 
    parameters $\theta_0$ and $\theta_1$ for increasing number of measurement shots. 
    As the number of shots increases, the landscape progressively smooths toward 
    the noiseless statevector limit $(n_{shots} \rightarrow \infty)$, illustrating the impact of finite sampling 
    noise on the optimization surface.}
    \label{fig:landscape_shots}
\end{figure}

In practice, we have access only to a finite sample of the probability distribution and must consider the effects of finite sampling on variational quantum algorithms. As illustrated in Figure \ref{fig:landscape_shots}, increasing the number of shots progressively smooths the optimization landscape — finite sampling introduces spurious local minima and produces a non-smooth surface, imposing additional challenges for gradient-based optimizers. The choice of CMA-ES \cite{grayver2016exploring} and COBYLA \cite{powell1994direct} was motivated by these considerations. In order not to exceed the regime where these algorithms perform worse than random search with replacement \cite{scriva2024challenges}, it is necessary to carefully balance the number of shots and iterations.

The surfaces in Figure \ref{fig:landscape_shots} were generated by randomly selecting 3 assets with seed 42 to compose a portfolio optimization instance in which exactly one asset is selected, using the ansatz

\section{Hardware experiments}
\label{appendix:c}

In order to assess the performance of the proposed ansatz on current noisy quantum hardware, we performed state preparation experiments and evaluated the Hamiltonian expectation value for each scenario. For these experiments, we selected the simulation run in which the CMA-ES optimizer yielded parameter vectors maximizing the sampling probability of the target bitstring $x^*$. In all experiments, the number of shots was set to 4096. Experiments were executed using the Sampler and Estimator primitives on the devices \texttt{ibm\_kingston}, \texttt{ibm\_fez}, and \texttt{ibm\_marrakesh}. The calibration data for all devices used in this study are available in the project repository (see Data Availability statement).

The similarity between the probability distributions obtained from simulation and hardware runs was evaluated using two complementary metrics. The first is the Kullback-Leibler divergence, defined as
\begin{eqnarray}
    D_{KL}(P \Vert Q) = \sum_{x \in \mathcal{X}}P(x)\log\frac{P(x)}{Q(x)}, \label{kl_div}
\end{eqnarray}
where $P$ and $Q$ are probabilities distributions. Note that $D_{KL}(P \Vert Q) = 0$ if and only if $P = Q$, and $D_{KL}(P \Vert Q) > 0$ indicates a discrepancy between the distributions, with larger values corresponding to greater divergence. It is also important to note that (\ref{kl_div}) is asymmetric, i.e. $D_{KL}(P \Vert Q) \neq D_{KL}(Q \Vert P)$ in general. The second metric is the Hellinger fidelity \cite{nielsen2010quantum}, defined as
\begin{eqnarray}
    F(P,Q) = \left(\sum_i\sqrt{p_iq_i}\right)^2,
\end{eqnarray}
where $0 \leq F(P, Q) \leq 1$. A value of $F(P,Q)=1$ indicates that $P = Q$, while $F(P, Q) = 0$ indicates that the distributions have disjoint support. Intermediate values reflect the degree of overlap between the two distributions.

The results obtained through the Sampler and Estimator primitives are presented in Tables \ref{tab:sampler} and \ref{tab:estimator}, respectively. The Sampler results reveal low fidelity and high KL divergence values in most cases, indicating that hardware noise significantly distorted the prepared distributions relative to the simulation reference. The Estimator results show relative errors of approximately $50\%$ across all QPUs and scenarios, confirming that the cumulative effect of two-qubit gate noise substantially degraded the objective function estimates. These results highlight the need for more advanced error mitigation techniques and circuit transpilation optimizations before the proposed ansatz can yield reliable results on current NISQ hardware.
\begin{table}[!htbp]
    \centering
    \caption{Hardware experiment results for each QPU and scenario, 
    reporting KL divergence, fidelity, and target bitstring probability.}
    \label{tab:hardware_results}
    \begin{tabular}{llccc}
        \toprule
        \textbf{QPU} & \textbf{Scenario} & \textbf{KL Divergence} & \textbf{Fidelity} & \textbf{Target Probability} \\
        \midrule
        \multirow{3}{*}{ibm\_kingston} 
            & I   & 5.540569 & 0.008845 & 0.000488 \\
            & II  & 0.452282 & 0.000940 & 0.000000 \\
            & III & 6.164810 & 0.003155 & 0.000732 \\
        \midrule
        \multirow{3}{*}{ibm\_fez} 
            & I   & 4.437095 & 0.008741 & 0.001953 \\
            & II  & 0.004935 & 0.000009 & 0.000000 \\
            & III & 7.079446 & 0.001906 & 0.000244 \\
        \midrule
        \multirow{3}{*}{ibm\_marrakesh} 
            & I   & 0.396861 & 0.002275 & 0.000000 \\
            & II  & 0.358606 & 0.000220 & 0.000000 \\
            & III & 6.500446 & 0.002231 & 0.000488 \\
        \bottomrule
    \end{tabular}
    \label{tab:sampler}
\end{table}

\begin{table}[!htbp]
    \centering
    \caption{Objective function values and relative errors obtained on real quantum 
    hardware for each QPU and scenario, where the relative error is defined as 
    $|\text{obj\_value} - \text{reference}|/|\text{reference}|$. The reference 
    values correspond to the objective function values obtained in simulation 
    experiments where the target bitstring $x^*$ was the most probable outcome 
    after training: $-0.8352$, $-0.8273$, and $-2.0527$ for Scenarios I, II, 
    and III, respectively.}
    \label{tab:hardware_obj_values}
    \begin{tabular}{llcc}
        \toprule
        \textbf{QPU} & \textbf{Scenario} & \textbf{Objective Value} & \textbf{Relative Error} \\
        \midrule
        \multirow{3}{*}{ibm\_kingston} 
            & Scenario I   & $-0.423282$ & $0.493171$ \\
            & Scenario II  & $-0.427963$ & $0.482729$ \\
            & Scenario III & $-0.980868$ & $0.522153$ \\
        \midrule
        \multirow{3}{*}{ibm\_fez} 
            & Scenario I   & $-0.412210$ & $0.506428$ \\
            & Scenario II  & $-0.424199$ & $0.487278$ \\
            & Scenario III & $-0.982028$ & $0.521587$ \\
        \midrule
        \multirow{3}{*}{ibm\_marrakesh} 
            & Scenario I   & $-0.437973$ & $0.475581$ \\
            & Scenario II  & $-0.421351$ & $0.490721$ \\
            & Scenario III & $-0.968446$ & $0.528204$ \\
        \bottomrule
    \end{tabular}
    \label{tab:estimator}
\end{table}

\end{document}